\documentclass[reprint, aps, pra, amsfonts, amsmath, twocolumn, superscriptaddress, 10pt, floatfix]{revtex4-2}
\usepackage{graphicx} 
\usepackage{natbib}
\usepackage[utf8]{inputenc}
\usepackage[english]{babel}
\usepackage{amsmath}
\usepackage{amsfonts}
\usepackage{amssymb}
\usepackage{xcolor}
\usepackage[T1]{fontenc}
\usepackage{mathrsfs}
\usepackage{braket}
\usepackage{geometry}
\usepackage{here}
\usepackage{lmodern}
\usepackage{array}
\usepackage{url}
\usepackage{textcase}
\usepackage{bm}
\usepackage{float}
\usepackage{caption} 
\usepackage{ragged2e}
\usepackage{float}

\begin{document}

 \title{Re-exploring Control Strategies in a Non-Markovian Open Quantum System by Reinforcement Learning}

\author{Amine Jaouadi}
 \affiliation{LyRIDS, ECE-Paris, Graduate School of Engineering, 75015 Paris, France}

\author{Etienne Mangaud}
 \affiliation{MSME, Universit\'e Gustave Eiffel, UPEC, CNRS, 77454 Marne-La-Vall\'ee, France }

\author{Mich\`ele Desouter-Lecomte}
 \email{michele.desouter-lecomte@universite-paris-saclay.fr}
\affiliation{Institut de Chimie Physique, Universit\'e Paris-Saclay-CNRS, UMR8000, 91400 Orsay, France}

\begin{abstract}
In this study, we reexamine a recent optimal control simulation targeting the preparation of a superposition of two excited electronic states in the UV range in a complex molecular system. We revisit this control from the perspective of reinforcement learning, offering an efficient alternative to conventional quantum control methods. The two excited states are addressable by orthogonal polarizations and their superposition corresponds to a right or left localization of the electronic density. The pulse duration spans tens of femtoseconds to prevent excitation of higher excited bright states what leads to a strong perturbation by the nuclear motions. We modify an open source software by L. Giannelli $\textit{et}$ $\textit{al}$., Phys. Lett. A, 434, 128054 (2022) that implements reinforcement learning with Lindblad dynamics, to introduce non-Markovianity of the surrounding either by time-dependent rates or more exactly by using the hierarchical equations of motion with the QuTiP-BoFiN package. This extension opens the way to wider applications for non-Markovian environments, in particular when the active system interacts with a highly structured noise.

\end{abstract}

\maketitle
	

\section{Introduction}
Since several decades, quantum-state manipulation with electromagnetic fields is a central problem in many areas of physics and chemistry to prepare particular initial states or realizing unitary transformations (quantum gates). The hardware systems and the spectral range are very different, from spins in nuclear magnetic resonance (NMR) \cite{Chuang2005} to molecular systems \cite{Vivie2018} or complex photosynthetic systems \cite{Plenio2012,Hoyer2014} and systems involved in emerging quantum technology \cite{Glaser2015,Sugny2022} for instance, superconducting quantum interference device (SQUID) \cite{Han2005}, trapped ions \cite{Monroe2004} or atoms \cite{Grangier2007}, nitrogen-vacancy diamond centers \cite{Avinadav2014} or quantum dots \cite{ Gross2008}. Quantum control has developed through various theoretical strategies as pump-dump schemes \cite{Tannor1985,Kosloff1989}, coherent control \cite{Brumer1992}, adiabatic methods \cite{Bergmann2017, Bergmann2019}, local \cite{Engel2009,Tavernelli2019} or Lyapunov control \cite{Petersen2012}, Pontryagin optimal control \cite{Boscain2021}, adaptative tracking \cite{ Rabitz2003} and optimal control theory (OCT) \cite{Rabitz1988} that involves a rich variety of optimization algorithms, for instance Rabitz monotonous convergent algorithm \cite {Rabitz1998, Turinici2003} or Krotov method \cite{Krotov1996, Palao2003}, gradient ascent pulse engineering GRAPE \cite{GRAPE2005,Glaser2011}, chopped random basis optimization CRAB \cite{CRAB2011,Qamar2019}.

Recently, there has been significant interest in applying reinforcement learning (RL), a distinctive machine learning technology, to quantum control and this begs the questions : Does RL suggest new control strategies ?  How does it compare with standard algorithms \cite{Zhang2019}?  What is the efficiency for control in dissipative environment \cite{An2021}? RL has already been applied in control for state preparation \cite{Bukov2018} and gate realization \cite{ An2019,Niu2019} and quantum compiling \cite{Prati2021} in quantum technology. Recently RL has recovered the well-known counter-intuitive STIRAP (Stimulated Raman Adiabatic Passage) pulse sequence in a three-state system \cite{Bergmann2017,Bergmann2019}. In reference \cite{Porotti2019} the laser may be on or off leading to a so-called digital-STIRAP ensuring an efficient transfer without the constraint of adiabaticity conditions \cite{Prati2020}. Conversely, in references \cite{Brown2021,Giannelli2022}, the pulses are continuous, while the RL algorithm optimizes either the laser detuning or the Rabi frequencies.  

In this work, our objective is to revisit with RL an optimal control simulation recently performed in a molecular system (phenylene ethynylene dimer) with $C_{2v}$ symmetry \cite{Jaouadi2022}. This benchmark case is interesting for different reasons. The system consists of two quasi-degenerate excited electronic states of different symmetries. They are addressable by orthogonal polarizations. The preparation of a superposition of the two states corresponds to a right or left localization of the electronic density in a way similar to  the localization in a double-well by superposing the two lowest eigenstates. Creating such a coherence in complex systems with orthogonal polarizations has been recently discussed  as a prospective avenue for achieving coherent control over excitonic energy transfer \cite{Kassal2019}.  In absence of dissipation, this is a V-type three-state system where two excited states are coupled to the ground state by two orthogonal transition dipoles. An analytical solution may be derived to prepare the superposition with equal weights \cite{Desouter2021}. This is an important landmark to test the control. On the other hand, the ideal electronic  V-type system strongly interacts with the surrounding leading to a non-Markovian non-perturbative open system.  The nuclear vibrations form two baths called the tuning and the coupling baths making fluctuate the energy gaps (also called longitudinal noise)  and the electronic coupling (transversal noise) respectively. 

In our previous OCT simulation, we employed hierarchical equations of motion (HEOM), which represent an exact method for addressing non-perturbative and non-Markovian open systems with Gaussian statistics \cite{Tanimura2020,Shi2018,Shi2021,Shi2023,Borrelli2021,mangaud2023survey}. In order to investigate the RL control, we start from the open source software \cite{RLpy} presented in reference \cite{Giannelli2022}. This software uses the libraries QuTip \cite{Qutip2012} and TensorFlow \cite{tensorflow2015}. The software already implements Lindblad dynamics with the QuTip collapse operators. In this work, we introduce non-Markovianity in different ways. In a simplified strategy, we first consider time-dependent rates \cite{Piilo2008,Anderson2014} by using time-dependent QuTip collapse operators. The rates are calibrated from the decoherence matrix \cite{Anderson2014} extracted from the exact HEOM dynamics \cite{MangaudMeier2017,MDL2018}. We then address the exact non-Markovian dynamics with the QuTip HEOMsolver \cite{Lambert2023}.  

The paper is organized as follows. In section \ref{sec:model} we describe the model treated as an isolated or an open system interacting with two Bosonic baths. We summarize the Lindblad and HEOM operational equations in section \ref{sec:Dynameth}. The control by RL or OCT is presented in section \ref{sec:control}. The RL results are given in section \ref{sec:results} and a comparison with OCT is made in section \ref{sec:compareRLOCT} before concluding in section \ref{sec:conclusion}.

\section{Model}
\label{sec:model}
\begin{figure}

\includegraphics[width =1.\columnwidth]{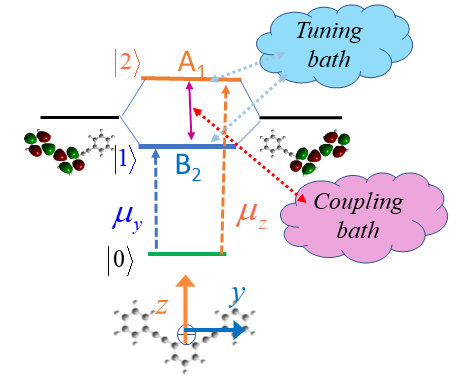}
\caption{\justifying{Schematic representation of the V-type model in 1,3-bis(phenylethynyl)benzene. The two excited states of symmetry $A_1$ and $B_2$ at the geometry of the ground state are delocalized over the two sites. Their superposition with equal weights corresponds to a localization on the left or right sites. The excited states are addressable by orthogonal dipole moments. The electronic sub-system is coupled to the vibrational baths. The tuning bath (longitudinal noise) making fluctuate the energy gap gathers the symmetric $A_1$ modes and the coupling bath (transversal noise) varying the electronic coupling contains the $B_2$ modes.}}  
\label{fig:fig1}
\end{figure}

The V-type three-state model of the dimer (1,3-bis(phenylethynyl)benzene) is schematized in Fig. \ref{fig:fig1}. It is calibrated from $\textit{ab}$ $\textit{initio}$ data computed by B. Lasorne $\textit{et}$ $\textit{al}$. \cite{Ho2017,Ho2019,Lasorne2023,Jaouadi2022} with the density-functional theory (DFT) for the ground state $S_0$ and the time-dependent density-functional theory (TDFT) for the two excited states ($S_1(B_2)$ and $S_2(A_1)$).  The energies of the two excited states ($E_{S_1}$ = 4.43 eV and $E_{S_2}$ = 4.47 eV) are taken at the equilibrium geometry of the ground state (planar with $C_{2v}$ symmetry). The two states are bright and coupled to the ground state by orthogonal transition dipoles. The axes are chosen so that $z (A_1)$ is the $C_2$ rotation axis, $y (B_2)$ lies within the molecular plane, and $x (B_1)$ is orthogonal to it. The respective transition dipoles are ${{\vec{\mu }}_{01}}=(0,{{\mu }_{y}},0)$ with $\mu_y = 3.96 ea_0$ and ${{\vec{\mu }}_{02}}=(0,0,{{\mu }_{z}})$ with $\mu_z = -1.83 ea_0$.

\subsection{Isolated system}
The system is defined by selecting the ground and the first two excited electronic states at the equilibrium geometry of the ground state (vertical Franck-Condon transition). It is an ideal system assumed to be frozen at this geometry. The two excited states are coupled by non-adiabatic interactions via a conical intersection \cite{Ho2019,Lasorne2023}. We choose a diabatic representation with states of $A_1$ or $B_2$ symmetry so that the electronic coupling vanishes at that reference position and the system Hamiltonian is simply:
\begin{equation}
H_{S}^{0}=\sum\nolimits_{j=0}^{2}{\left| j \right\rangle }\left\langle  j \right|
\end{equation}

 The electronic coupling (between the two diabatic excited states) becomes different from zero when any vibration of $B_2$ symmetry is active. The two bright states are coupled to the ground state only radiatively and the time-dependent Hamiltonian at the dipolar approximation is :
\begin{equation}
H_S(t)={{H}_S^0}-\sum\limits_{j=1}^{2}{\left( \vec{{\mu} }_{0j}\vec{\mathcal{E}_{j}}(t)\left| 0 \right\rangle \left\langle  j \right|+hc \right)}
\label{eq:Hdet}
\end{equation}
where $hc$ designates the hermitian conjugate. The two fields are linearly polarized with ${{\vec{\mathcal{E}}}_{1}}=(0,{\mathcal{E}_{y}},0)$ and ${{\vec{\mathcal{E}}}_{2}}=(0,0,{\mathcal{E}_{z}})$. In interaction representation (I) and with the rotating wave approximation (RWA) \cite{Fujii2017} the Hamiltonian becomes :
\begin{equation}
\mathbf{H}_{S,I}^{RWA}(t)=-\frac{\hbar }{2}\left( \begin{matrix}
   0 & {{\Omega }_{y}}(t) & {{\Omega }_{z}}(t)  \\
   {{\Omega }_{y}}(t) & -2{{\Delta }_{y}} & 0  \\
   {{\Omega }_{z}}(t) & 0 & -2{{\Delta }_{z}}  \\
\end{matrix} \right)
\end{equation}
where the Rabi frequencies are ${{\Omega }_{y}}(t)={{\mu }_{y}}{{E}_{y}}(t)/\hbar$, ${{\Omega }_{z}}(t)={{\mu }_{z}}{{E}_{z}}(t)/\hbar$ and ${{E}_{j}}(t)$ ($j = y,z)$ are the pulse envelopes. $\Delta_{y}$ and $\Delta _{z}$ are the field detunings. 

\begin{figure}
\includegraphics[width =1.\columnwidth]{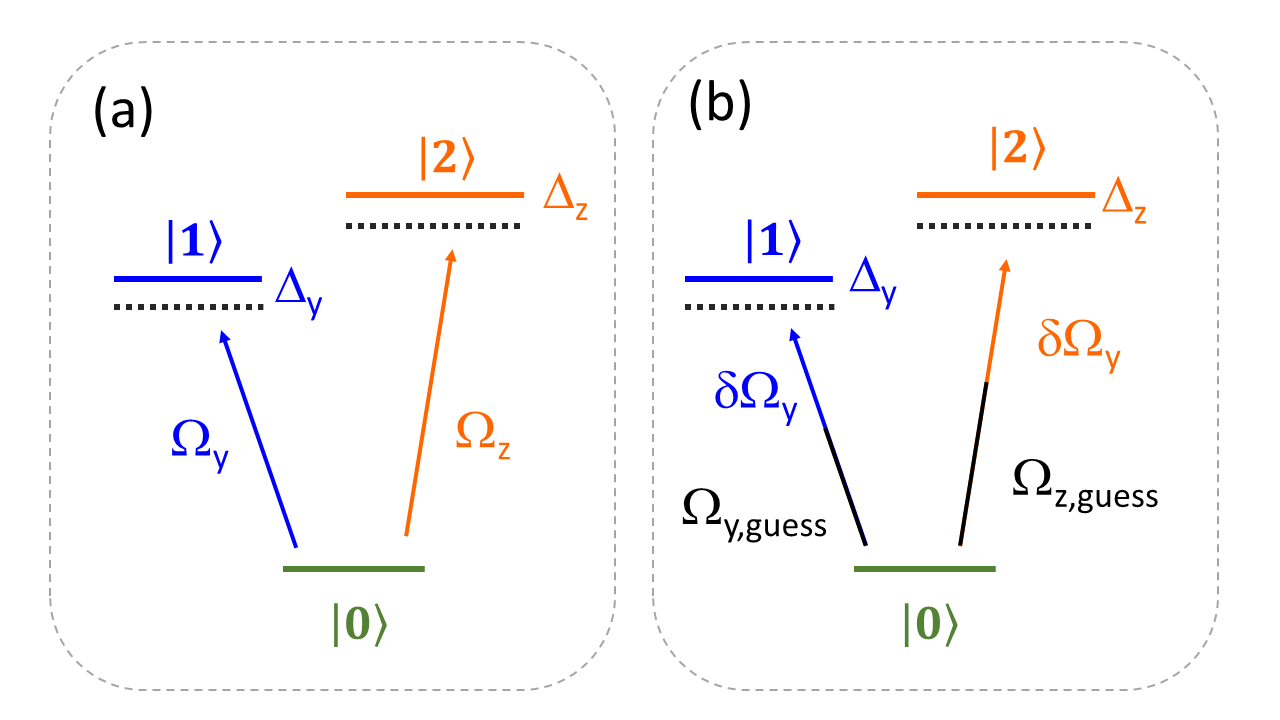}
\caption{\justifying{Actions optimized by the RL algorithm to prepare the target superposed state in the V-type system. (a) The actions are the Rabi frequencies $\Omega_y$ and $\Omega_z$; (b) the actions are the variations of the Rabi frequencies with respect to guess fields. The laser detuning $\Delta_y$ and $\Delta_z$ are assumed to vanish in this application. } }  
\label{fig:fig2}
\end{figure}

The two Rabi frequencies or their variations with respect to a guess field are the actions that will be optimized by the RL algorithm. They are represented in Fig. \ref{fig:fig2}. In our application, the target is the superposition of the two excited states with equal weights
\begin{equation}
 \left| 0 \right\rangle \to \frac{1}{\sqrt{2}}\left( \left| 1 \right\rangle +\left| 2 \right\rangle  \right).
\label{eq:target}
\end{equation}
This target is different from the superposition of the initial state and one excited state that may be prepared by fractional-STIRAP \cite{Bergmann2017} or by a $\pi/2$ pulse.  

By imposing equal Rabi frequencies at all times, i.e. pulses of the same duration $T$ with amplitudes in the inverse ratio of the dipole moments, the target transition is realized if each area is equal to $\pi /\sqrt{2}$ \cite{Desouter2021}
\begin{equation}
\int_{0}^{T}{\Omega_j (t)dt=\pi /\sqrt{2}}
\label{eq:pirule}
\end{equation}
with $j = y,z$. It is a generalization of the well-known $\pi$ rule for complete population transfer \cite{Thomas1983} or $\pi/2$ for creating a superposition involving the initial state.

\subsection{Open quantum system}
\begin{figure}
    \includegraphics[width =1.\columnwidth]{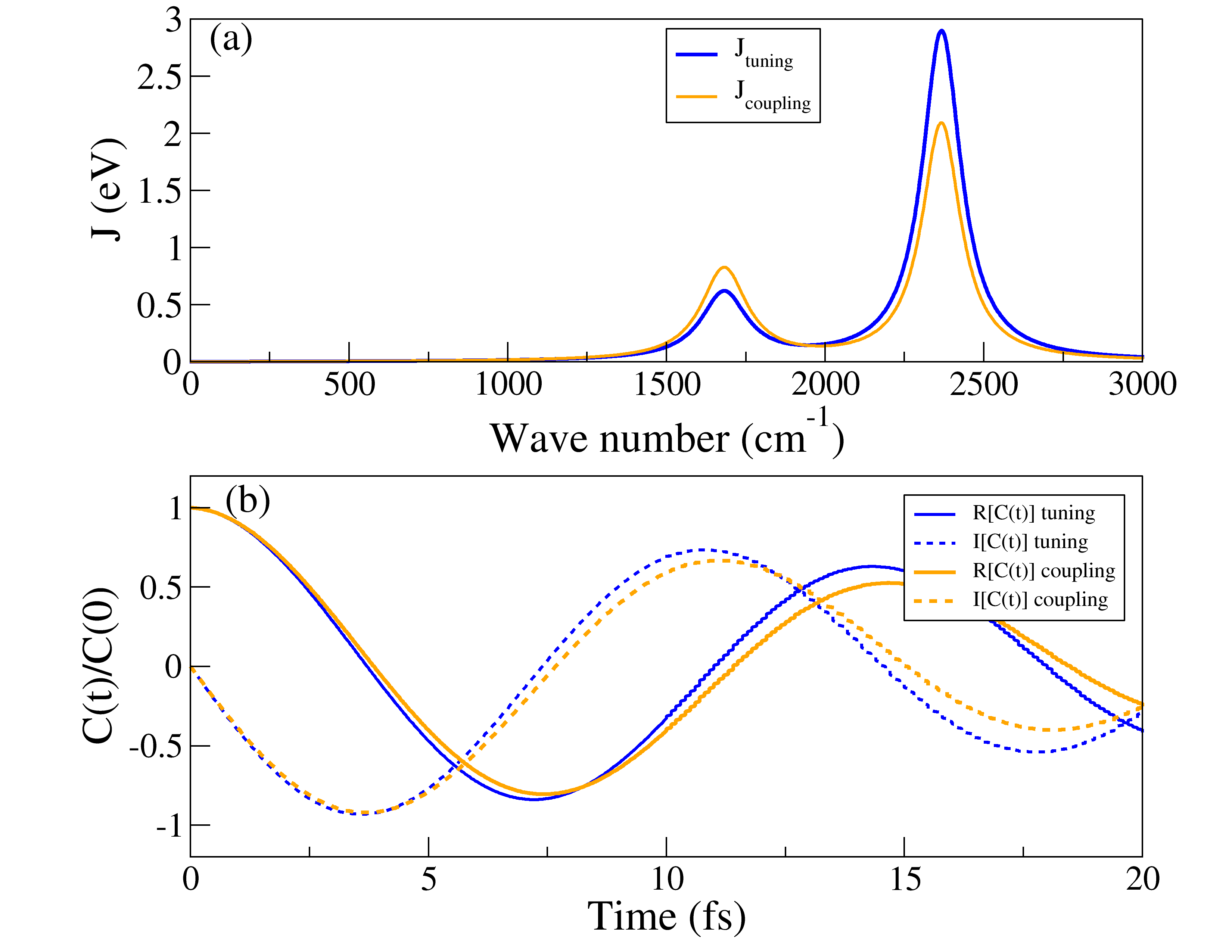}
    \caption{\justifying{(a) Spectral densities giving the strength of the system-bath coupling to the diagonal ($J_{tuning}$) or off-diagonal ($J_{coupling}= J_{S_1S_2}$) element of the system Hamiltonian block related to the two excited states. $J_{tuning}$ is $J_{S_1}$ and $J_{S_2}$ is assumed to be similar with $J_{S_2}= 0.797 J_{S_1}$ \cite{Jaouadi2022}. The tuning modes are of $A_1$ symmetry and the coupling ones of $B_2$ symmetry. (b) Real and imaginary parts of the normalized bath correlation function $C(t)$ (Eq.(\ref{Cdet})) at room temperature (zoom on the early 20 fs timescale). $C(t)$ decays in 200 fs.} }
    \label{fig:fig3}
\end{figure}

According to the chosen partition, all the nuclear vibrations belong to the baths. The $A_1$ modes make fluctuate the energies and the $B_2$ vibrations modify the electronic coupling that becomes different from zero when the $C_{2v}$ symmetry is broken. The generic Hamiltonian of the system-bath partition is then written
\begin{equation}
H(t)=H_S(t)+H_{SB}+H_B    
\end{equation}

where $H_B$ is the ensemble of the vibrational modes assumed to be harmonic. The normal modes are assumed to be the same in each electronic states but their equilibrium positions differ. $H_{SB}$ is the linear system-bath coupling. The two groups of $A_1$ or $B_2$ modes then constitute different baths that may be called the tuning baths coupled to the diagonal elements $\left| 1 \right\rangle \left\langle  1 \right|$ and $\left| 2 \right\rangle \left\langle  2 \right|$ of the system Hamiltonian and the coupling bath coupled to the off-diagonal elements $\left| 1 \right\rangle \left\langle  2 \right|$ and $\left| 2 \right\rangle \left\langle  1 \right|$ between the two excited states. This kind of partition in the case of a conical intersection has been discussed in different works applying HEOM dynamics \cite{Thorwart2016,Desouter2019,Desouter2021,Jaouadi2022}. The partition leads to a strong system-bath coupling and a non-Markovian master equation. The analysis of the dimer model is given in our previous work \cite{Jaouadi2022} where is explained how are obtained the continuous spectral densities $J(\omega)$  for the tuning and coupling baths from $\textit{ab}$ $\textit{initio}$ data, i.e. from the energy gradients and gradient of the electronic coupling at the reference position. The spectral densities give the strength of the coupling to the system for each energy range of the baths.  We select the main part of the spectral densities  consisting in very sharp peaks around 1700 and 2300 cm$^{-1}$. The two spectral densities $J_{tuning}(\omega)$ and $J_{coupling}(\omega)$ are presented in Fig. \ref{fig:fig3}(a). Figure \ref{fig:fig3}(b) gives the corresponding correlation functions of the collective mode of each bath 
\begin{equation}
C\left( t \right)=\frac{1}{\pi }\int\limits_{-\infty }^{+\infty }{d\omega \frac{J\left( \omega  \right){{e}^{i\omega \left( t \right)}}}{{{e}^{\beta \omega }}-1}} 
\label{Cdet}
\end{equation}
where $\beta$ is the Boltzmann constant. Due to the peaked shape of the spectral densities, the effective collective modes are under-damped and decay in about 200 fs. We zoom on the early time range of 20 fs, which is the pulse duration chosen in our simulations. Indeed, the pulse duration must be longer than about 10 fs to correspond to a sufficiently narrow spectral range to avoid the excitation of higher bright states. It is noticeable that the collective bath modes undergo a full oscillation within the 20 fs timescale. This serves as an indicator of non-Markovian behavior, as we will delve into further in the subsequent discussion. 

\section{Dynamical methods}
\label{sec:Dynameth}
The dynamics of open quantum systems \cite{Breuer2002} has been reviewed, as seen in references \cite{deVega2017}, and \cite{Koch2016} specifically focusing on control aspects. The applications of RL control \cite{Giannelli2022,An2021} usually assume a weak coupling and a Markovian bath treated by a Lindblad master equation \cite{Lindblad1976,Manzano2020}. We first summarize the main relations to introduce time-dependent rates and then we recall the operational equations for HEOM.

\subsection{Lindblad master equation}
\label{subsec:Lindblad}
For a $N-$dimensional system coupled to $M$ dissipative channels, the generic Lindblad operator reads:
\begin{equation}
\mathcal{D}(t )=\sum\limits_{k,l}^{M}{{{\Gamma }_{kl}}}\left( {{L}_{k}}\rho_t L_{l}^{\dagger }-\frac{1}{2}\left\{ L_{l}^{\dagger }{{L}_{k}},\rho_t  \right\} \right)    
\end{equation}
where  $\left\{ A,B \right\}=AB+BA$ denotes the anticommutator and the rates $\Gamma_{kl}$ are constant. Often, only the main dissipative processes are retained, as the radiative decay towards a sink \cite{Giannelli2022}, or dephasing processes affecting the diagonal elements of the density matrix or relaxation inducing population transfer due to inter-state coupling \cite{An2021}. In this work, we will consider these two processes induced by the tuning and coupling baths respectively leading to four operators ${{L}_{1}}=\left| 1 \right\rangle \left\langle  1 \right|$, ${{L}_{2}}=\left| 2 \right\rangle \left\langle  2 \right|$, ${{L}_{3}}=\left| 1 \right\rangle \left\langle  2 \right|$ and ${{L}_{4}}=\left| 2 \right\rangle \left\langle  1 \right|$ (they are the collapse operators in QuTip \cite{Qutip2012}). Moreover, we want to account for non-Markovian baths, at least on an approximate way by introducing time-dependent rates \cite{Piilo2008,Anderson2014}. This has given rise to many fundamental analysis 
\cite{Anderson2014,Rivas2014,Tanimura2017,MDL2018}
 concerning the non-markovianity signature or the positivity of the dynamical map \cite{Sugny2023}. Indeed, the master equation of non-Markovian dynamics can be recast in a Lindblad form with time-dependent rates that may be transitory negative. If the Lindblad dissipator is expressed with the orthogonal basis set of ${{N}^{2}}$ operators formed by the normalized identity $G_0=I/\sqrt{N}$ and the $N^2-1$ generators of $SU(N)$, $G_i (i = 1,...,N^2-1)$ \cite{Kimura2003,Aerts2014}, which are the Pauli matrices for $N = 2$ and the Gell-Mann matrices for $N = 3$, the corresponding rate matrix is also called the time dependent decoherence matrix $D_{jk}$ \cite{Anderson2014}
\begin{equation}
\mathcal{D}(t)=\sum\limits_{j,k=1}^{{{N}^{2}}-1}{{{D}_{jk}}(t)}\left( {{G}_{j}}\rho_t {{G}_{k}}-\frac{1}{2}\left\{ {{G}_{k}}{{G}_{j}},\rho_t  \right\} \right).
\label{eq:matdeco}
\end{equation}
The eigenvalues are the canonical decay rate $\Gamma^c_k$ associated to the time-dependent decay channels. The decoherence matrix is given by \cite{Anderson2014} 
\begin{equation}
{{D}_{ij}}(t)=\sum\nolimits_{m=1}^{{{N}^{2}}-1}{Tr\left[ {{G}_{m}}{{G}_{i}}{{\Lambda }_{t}}\left[ {{G}_{m}(t)} \right]{{G}_{j}} \right]}
\end{equation}  
where $\Lambda_{t}\left[ .\right]$ denotes the map of the time local non-Markovian master equation ${{\dot{G}}_{m}}(t)={{\Lambda }_{t}}\left[ {{G}_{m}(t)} \right]$ \cite{Hall2007,Anderson2014}. This requires $(N^2-1)$ propagations of the basis operators performed here with HEOM as shown in \cite{MangaudMeier2017,MDL2018}. Some elements of the decoherence matrix will be used to calibrate the time-dependent rates of the selected collapse operators. 

This is a low-cost way to introduce easily some non-Markovianity with time-dependent collapse operators. However, its efficiency might be somewhat limited, as the rates are calibrated based on field-free dynamics and are subject to potential modification by the applied fields \cite{MDL2018}. 

\subsection{HEOM}
\label{subsec:HEOM}
We now summarize the main operational equations of the HEOM method. The system density matrix is the partial trace of the full density matrix ${{\rho }_{tot}}(t)$ over the bath degrees of freedom $\rho (t)=T{{r}_{B}}\left[ {{\rho }_{tot}}(t) \right]$. The initial condition is assumed to be factorized ${{\rho }_{tot}}(0)=\rho (0){{\rho }_{eq}}$ where ${\rho }_{eq}$ is the density matrix of the baths at Boltzmann equilibrium at a given temperature. The HEOM may be considered as a numerically exact method for non-Markovian dynamics with harmonic baths when convergence is achieved by a relevant truncation of the hierarchy. The method is abundantly described in the literature, see for instance references \cite{Tanimura2020} for a recent review or \cite{mangaud2023survey} for a pedagogical survey, \cite{Shi2018,Shi2021,Shi2023,Borrelli2021} for applications with the tensor-train format and \cite{Lambert2023} for a review of different softwares, in particular that implemented in QuTip. We briefly recall that the master equation is solved by a time local system of coupled equations among auxiliary density matrices or auxiliary density operators (ADOs) arranged in a hierarchical structure. The algorithm requires a particular fit of the correlation function $C(t)$ as a sum of $K$ damped oscillatory terms also called artificial decaying modes 
\begin{equation}
C\left( t \right)=\sum\limits_{k=1}^{K}{{{\alpha }_{k}}{{e}^{i{{\gamma }_{k}}t}}} 
\label{eq:C(t)alpha}
\end{equation}
and ${{C}^{*}}\left( t \right)=\sum\limits_{k=1}^{K}{{{{\tilde{\alpha }}}_{k}}{{e}^{i{{\gamma }_{k}}t}}}$. Analytical expressions for the ${{\alpha }_{k}}$, ${\tilde{\alpha }}_{k}$ and ${\gamma }_{k}$ parameters can be derived from Eq.(\ref{Cdet}) \cite{Tannor2010} when the spectral density is fitted by a sum of two-poles Tannor-Meier Lorentzian functions \cite{Meier1999}
\begin{equation}
J\left( \omega  \right)=\sum\limits_{l=1}^{{{n}_{l}}}{\frac{{{p}_{l}}\omega }{\left[ {{\left( \omega +{{\Omega }_{l}} \right)}^{2}}+\Gamma _{l}^{2} \right]\left[ {{\left( \omega -{{\Omega }_{l}} \right)}^{2}}+\Gamma _{l}^{2} \right]}}.
\label{eq:two_polefunction}
\end{equation}
The parameters fitting the spectral densities of Fig. \ref{fig:fig3} are given in reference \cite{Jaouadi2022}. Each ADO is labelled by a collective index $\mathbf{n}=\left\{ {{n}_{1}},\cdots ,{{n}_{K}} \right\}$ specifying the occupation number of each artificial mode. The system density matrix has the index $\mathbf{n}=\left\{ 0,\cdots ,0 \right\}$. The HEOM equations are:
\begin{align}
  & {{{\dot{\rho }}}_{\mathbf{n}}}(t)={{L}_{S}(t)}{{\rho }_{\mathbf{n}}}(t)+i\sum\limits_{k=1}^{K}{{{n}_{k}}{{\gamma }_{k}}{{\rho }_{\mathbf{n}}}}(t) \nonumber \\
  &-i\left[ S,\sum\limits_{k=1}^{K}{{{\rho }_{\mathbf{n}_{k}^{+}}}(t)} \right]  \nonumber \\
 & -i\sum\limits_{k=1}^{K}{{{n}_{k}}\left( {{\alpha }_{k}}{{S}}{{\rho }_{\mathbf{n}_{k}^{-}}}(t)-{{{\tilde{\alpha }}}_{k}}{{\rho }_{\mathbf{n}_{k}^{-}}}(t)S \right)}
\label{align:HEOM} 
\end{align}
where $L_S(t)$ is the system Liouvillian and $\mathbf{n}_{k}^{+}=\left\{ {{n}_{1}},\cdots ,{{n}_{k}}+1,\ldots ,{{n}_{K}} \right\}$, and  $\mathbf{n}_{k}^{-}=\left\{ {{n}_{1}},\cdots ,{{n}_{k}}-1,\ldots ,{{n}_{K}} \right\}$.

The OCT simulations make use of the HEOM (Eqs. (\ref{align:HEOM})) via our in-house developed software \cite{Jaouadi2022}. In RL simulations \cite{Giannelli2022} based on QuTip software, we use the QuTip-BOFiN HEOMsolver \cite{Lambert2023} that allows the description of the Bosonic baths by giving the real and imaginary parts of the correlation function $C(t)=C_R(t) +i C_I(t)$. For each bath, they are parametrized by a combination of decaying terms ${{C}_{R}}(t)=\sum\limits_{k=1}^{{{N}_{R}}}{c_{k}^{R}}{{e}^{-\gamma _{_{k}}^{R}t}}$ and ${{C}_{I}}(t)=\sum\limits_{k=1}^{{{N}_{I}}}{c_{k}^{I}}{{e}^{-\gamma _{_{k}}^{I}t}}$ where the $c_k$ and $\gamma_k^{R,I}$ are complex. This is an alternative to expansion of Eq.(\ref{eq:C(t)alpha}) already adopted for the second order  time non-local \cite{Meier1999} or time-local non-Markovian equations \cite{Kleinekathöfer2004}. The HEOM equations adapted to this partition of the correlation function in real and imaginary part are given in Eq.(11) of reference \cite{Lambert2023}. The analytical expressions of the  $c_k$ and $\gamma_k^{R,I}$ parameters when the spectral density is fitted by the two-pole Lorentzian functions (Eq.(\ref{eq:two_polefunction})) are given in references \cite{Meier1999,Kleinekathöfer2004}. 

\section{Control} 
\label{sec:control}
\subsection{Reinforcement learning}
\begin{figure}
    \includegraphics[width =1.\columnwidth]{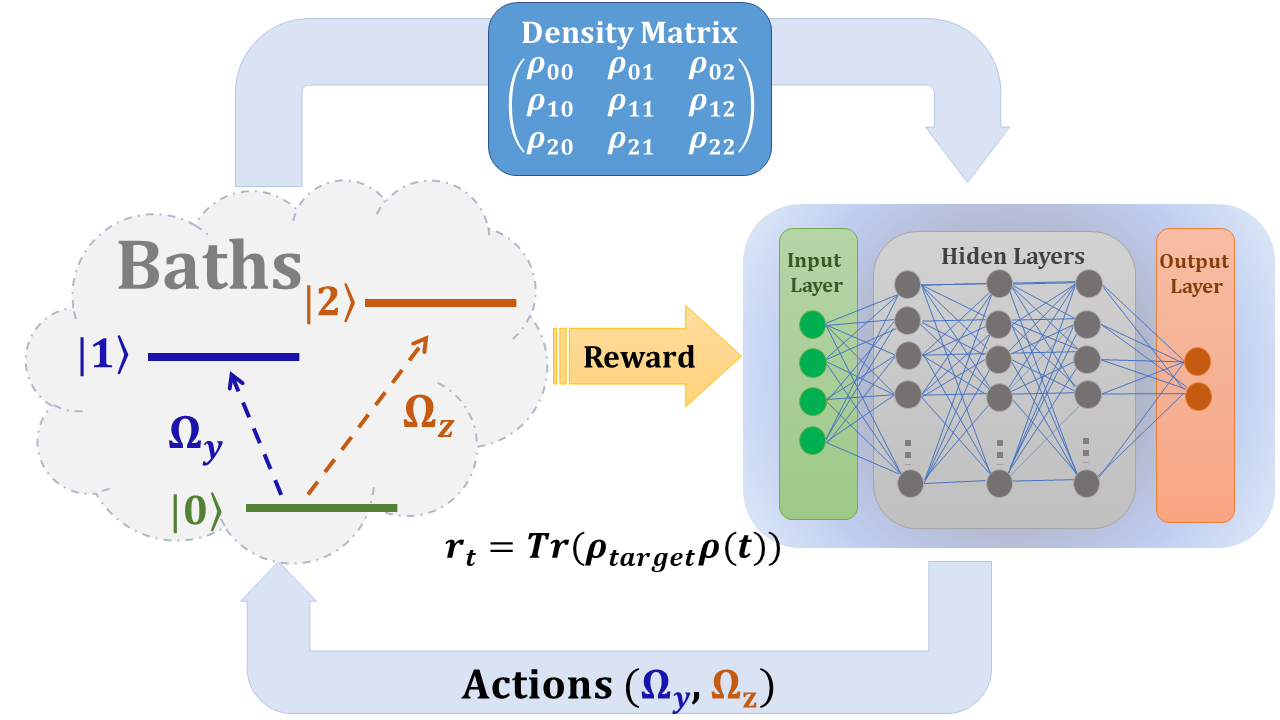}
    \caption{\justifying{Representaion of a cycle of the RL. At each time, the reduced density matrix of the three-state system coupled to its surrounding is the input of the neural network. The policy $\pi \left( {{a}_{t}}\left| {{s}_{t}} \right. \right)$ is optimized and provides two actions that are the pulse Rabi frequencies. The reward is the fidelity to reach the target, which is here the superposed state of the two excited states.}}
    \label{fig:fig4}
\end{figure}

The RL algorithm is summarized in many references, for instance \cite{Sutton2018,Carleo2019}.  By using the generic vocabulary, the principle is as follows. A target must be reached in an environment. At each time, an agent makes an observation and gets information about its state. The agent then chooses an action according to a policy to modify the state. The agent obtains a reward that estimates the progress towards the target. In our application, we have thus to define the environment, the agent, the action, and the reward. The four points are represented in Fig. \ref{fig:fig4}. The RL environment is the active system and its surrounding, i.e.,  the V-three-level system coupled to both tuning and coupling baths.  The observation is the state of the system described by the reduced density matrix solution of a master equation. The agent is an algorithm called REINFORCE \cite{Sutton2018}. It uses a neural network with three hidden layers in our application. The input layer contains all the density matrix elements (nine in our case). The output layer may provide discrete values or a continuous distribution. The outputs are discrete when only some actions are available, for instance if the laser may be only on or off, giving four outcomes in the two-pulse case \cite{Prati2020}. When the distribution is continuous, the output layer gives parameters of this distribution, for instance the mean of Gaussian distributions for the actions ${{a}_{y,z}}$ that provide the Rabi frequencies ${{\Omega }_{y}}$ and ${{\Omega }_{z}}$ of the two pulses. The reward is the control fidelity ${{r}_{t}}=Tr(\rho_ {target}^\dagger\rho(t))$.  The policy is the conditional probability $\pi \left( {{a}_{t}}\left| {{s}_{t}} \right. \right)$ that the agent takes action $a$ when the system state is $s$. The action at a time $t$ only depends on the state at that time and the process is called a Markov decision chain. Note that this does not mean that the dynamics of the system must be Markovian. The Markov decision chain means that when two states $s$ and $s'$ observed by the agent are the same, the probability to choose $a$ is the same regardless of the history to reach the state $s$.

We employ the policy gradient method \cite{Williams1992}, which is a technique used in reinforcement learning. It involves adjusting the policy's parameters aiming to maximize the cumulative reward over time. All the parameters of the network are represented by the global index $\theta$. They are initialized at random. One then generates a batch of $M$ episodes with the current policy $\pi \left( {{a}_{t}}\left| {{s}_{t}} \right. \right)$. An episode or trajectory $\tau$ is divided in $N$ time steps and lasts $T =N \delta t$. For each episode, one collects the $N$ data triplets $(s_t, a_t, r_t )$ where $t = i \delta t$ for the $i^{th}$ time step. The performance of the agent is estimated by the so-called return $R$ that depends on the network parameters $\theta$ and is the main tool to optimize the policy. For each episode, the return $R(\tau)$ may be defined in more or less sophisticated ways by the simple sum of all the rewards $r_t$ or a weighted sum of these with a discount rate \cite{Bukov2018,Lamata2021,An2021}. Here, $R(\tau)$ is the sum of the $r_t$ and all the $r_t = 0$ if $ t < T$ so that $R(\tau) = r_N$, i.e. it is given by the final control fidelity. For the bunch of $M$ episodes driven by the same policy, the return is the expectation value 
\begin{equation}
E\left[ R \right]=\sum\limits_{\tau =1}^{M}{{{p}_{\theta }}(\tau )R(\tau )} 
\label{eq:expectreturn}
\end{equation}
where ${p}_{\theta }(\tau )$ is the probability of driving trajectory $\tau$. Following reference \cite{Marquardt2021} we summarize the main points of the policy optimization. The probability of each trajectory is different since the actions are taken at random in the current policy. It is a product for each time step of the probability $p({{s}_{t+1}}\left| {{a}_{t}},{{s}_{t}} \right)$ to have a transition from state $s_t$ to state $s_{t+1}$  induced by action $a_t$ times the probability for the agent to choose action $a_t$ for state $s_t$ :  
\begin{equation}
{{p}_{\theta }}(\tau )=\prod\limits_{t}^{N}{p({{s}_{t+1}}\left| {{a}_{t}},{{s}_{t}} \right.}){{\pi }_{\theta }}({{a}_{t}},{{s}_{t}}).    
\end{equation}
$p({{s}_{t+1}}\left| {{a}_{t}},{{s}_{t}} \right)$ does not depend on the parameters $\theta$ but only on the system dynamics. Therefore, the gradient of the average return involves only the gradient of the policy 
\begin{align}
  & {{\nabla }_{\theta }}{{p}_{\theta }}(\tau ) \nonumber \\    
 & =\sum\limits_{t=1}^{N}{\prod\limits_{t'=1}^{N}{p(\left. {{s}_{t'+1}} \right|}}{{a}_{t'}},{{s}_{t'}}){{\pi }_{\theta }}({{a}_{t'}},{{s}_{t'}}) \nonumber \\  
 & \quad \times {{\nabla }_{\theta }}\ln {{\pi }_{\theta }}({{a}_{t}},{{s}_{t}})  \nonumber \\  
 & ={{p}_{\theta }}(\tau )\sum\limits_{t=1}^{N}{{{\nabla }_{\theta }}\ln {{\pi }_{\theta }}({{a}_{t}},{{s}_{t}})}. 
\label{align:gradptheta}
 \end{align}
The network parameters are optimized so that their gradient $\nabla \theta$ is parallel to the gradient of the average return with a factor $\eta$ called the learning rate
$\nabla \theta =+\eta {{\nabla }_{\theta }}E\left[ R \right]$
and by Eqs.
(\ref{eq:expectreturn}) and (\ref{align:gradptheta}) one has
\begin{align}
  & \nabla \theta =+\eta \sum\limits_{\tau =1}^{M}{{{p}_{\theta }}(\tau )R(\tau )\sum\limits_{t=1}^{N}{{{\nabla }_{\theta }}}}\ln {{\pi }_{\theta }}({{a}_{t}},{{s}_{t}}) \nonumber \\ 
 & =+\eta E\left[ R(\tau )\sum\limits_{t=1}^{N}{{{\nabla }_{\theta }}\ln {{\pi }_{\theta }}({{a}_{t}},{{s}_{t}})} \right]. 
\end{align}
The parameters are updated according to the logarithmic gradient of the policy times the return and the learning rate that must be chosen not too fast and not too slow. Since the gradient contains the return, all the actions become more likely, the more the return is larger. 
An important point is that the optimization algorithm of the network parameters is independent of the underlying dynamical model. This is a difference with the standard optimization in OCT where the gradient of the final fidelity involves the system Hamiltonian. RL operates beyond static databases; it collects data during training.

\subsection{Optimal control}
The optimal field is built by iterations to maximize the cost functional that is the fidelity $\mathcal{F}=Tr\left[ {{\rho }_{target}^\dagger\rho (T)} \right]$ at the final time $T$ with constraints to restrain the field intensity
and to fulfill the master equation at any time. The optimization is performed here by Rabitz' monotonously convergent algorithm \cite{Rabitz1998,Ohtsuki1999} that involves forward and backward propagation of the system density matrices with initial condition $\rho(0)$ and of an auxiliary system density matrix $\chi(t)$ with final condition $\chi(T) = \rho_{target}$. It is worth noting that a two-point boundary-value quantum control paradigm (TBQCP) has been presented in the literature as an accelerated convergent algorithm \cite{HoRabitz2010}. However for the sake of simplicity this method is not used in our simulations. Dynamics is driven  with HEOM. The iterations begin with a guess field that strongly influences the final field. The operational relations for the backward propagation are given in references \cite{MDL2018,Jaouadi2022}. The field at each iteration $k$ is obtained by ${{\varepsilon }^{(k)}}={{\varepsilon }^{(k-1)}}+\Delta {{\varepsilon }^{(k)}}$ where $\Delta {{\varepsilon }^{(k)}}$ is estimated by 
\begin{equation}
\Delta \varepsilon (t)=\frac{1}{\alpha }\Im m\left\{ Tr\left( \chi (t)\left[ \sum\nolimits_{p}{{{\mu }_{p}}},\rho (t) \right] \right) \right\} 
\label{eq:champOCT} 
\end{equation}
where $\alpha$ is the intensity penalty factor. 
Note that we do not use RWA in this approach.

\section{Control by RL}
\label{sec:results}
 We choose a pulse duration of $T = 20$ fs. This is relevant to avoid a too large spectral band that would imply higher bright excited states not included in the model system. For RL simulations, the laser detunings are assumed to vanish, so the only optimized parameters are the two Rabi frequencies $\Omega_y$ and $\Omega_z$ at all times. In all the RL examples, the results are given in reduced units for the time ($t/T$) and the Rabi frequency ($T\Omega$). Conventional units are used in some OCT examples. 

 The three hidden network layers contain 100, 50 and 30 neurons. The learning rate has its standard value $\eta = 10^{-3}$. Our investigation has confirmed the critical significance of these meta-parameters. Reducing the number of neurons results in a decelerated convergence rate, while elevating it prolongs computational duration without commensurate convergence enhancement. Increasing the learning rate by a factor 10 is not efficient to reach the desired target.  Data are collected during a bunch of $M = 10$ episodes that are divided in 50 time steps. 

\subsection{RL in the isolated system}
\begin{figure}
    \includegraphics[width =1.\columnwidth]{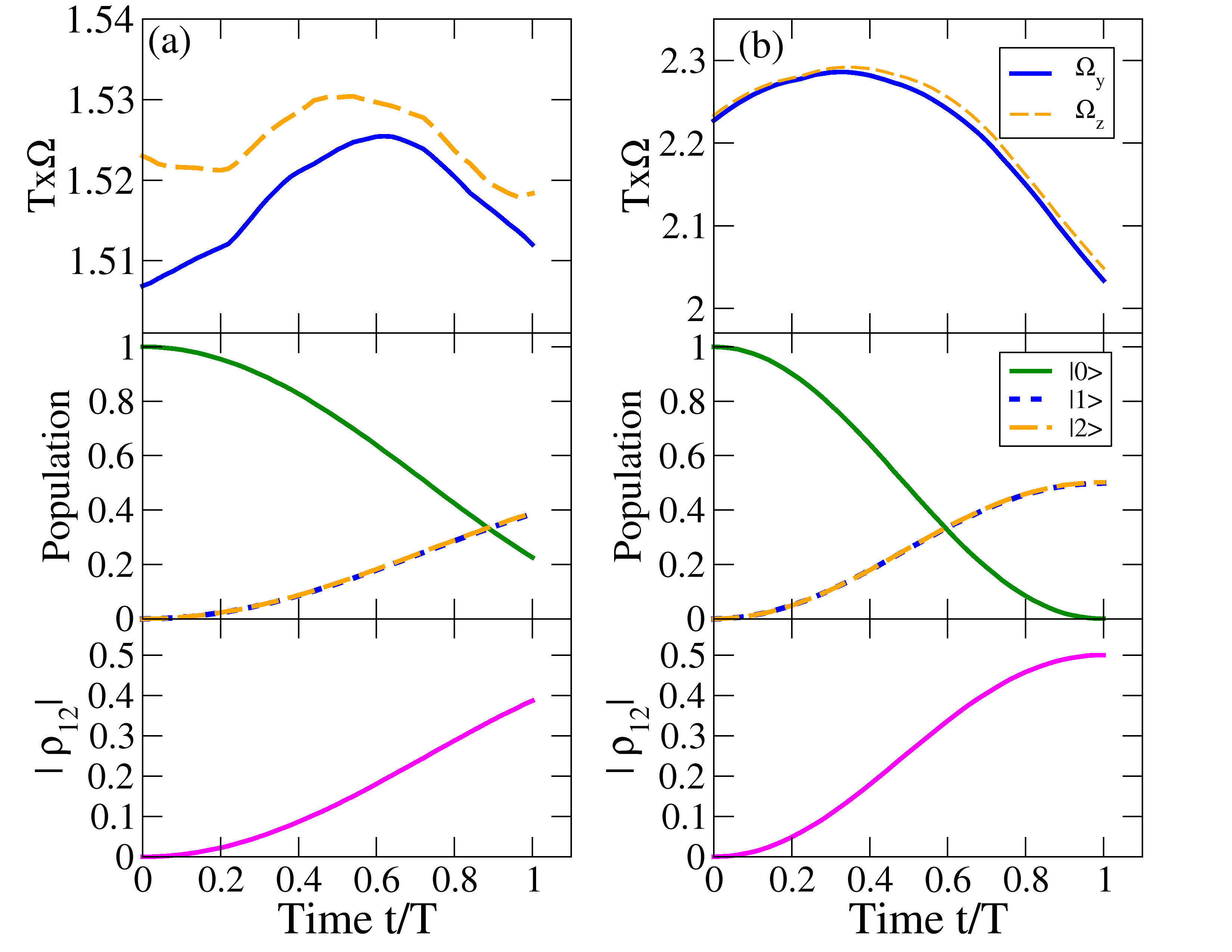}
    \caption{\justifying{Optimization of the superposed target state (Eq.(\ref{eq:target})) by RL in the isolated V-three-level system and $T\Omega_{max}=3$. The upper panels give the Rabi frequencies in reduced units ($T\Omega$), the middle ones show the populations in each state and the lower ones, the modulus of the coherence $\rho_{12}$ between the two excited states. (a) First iteration with initial random Rabi frequencies. The integrated frequencies are 1.54 for $\Omega_y$ and 1.55 for $\Omega_z$. (b) After 100 episodes. Both areas are 2.22 ($\pi / \sqrt{2}$), the best expected result.}  }
    \label{fig:fig5}
\end{figure}

\begin{figure}
    \includegraphics[width =1.\columnwidth]{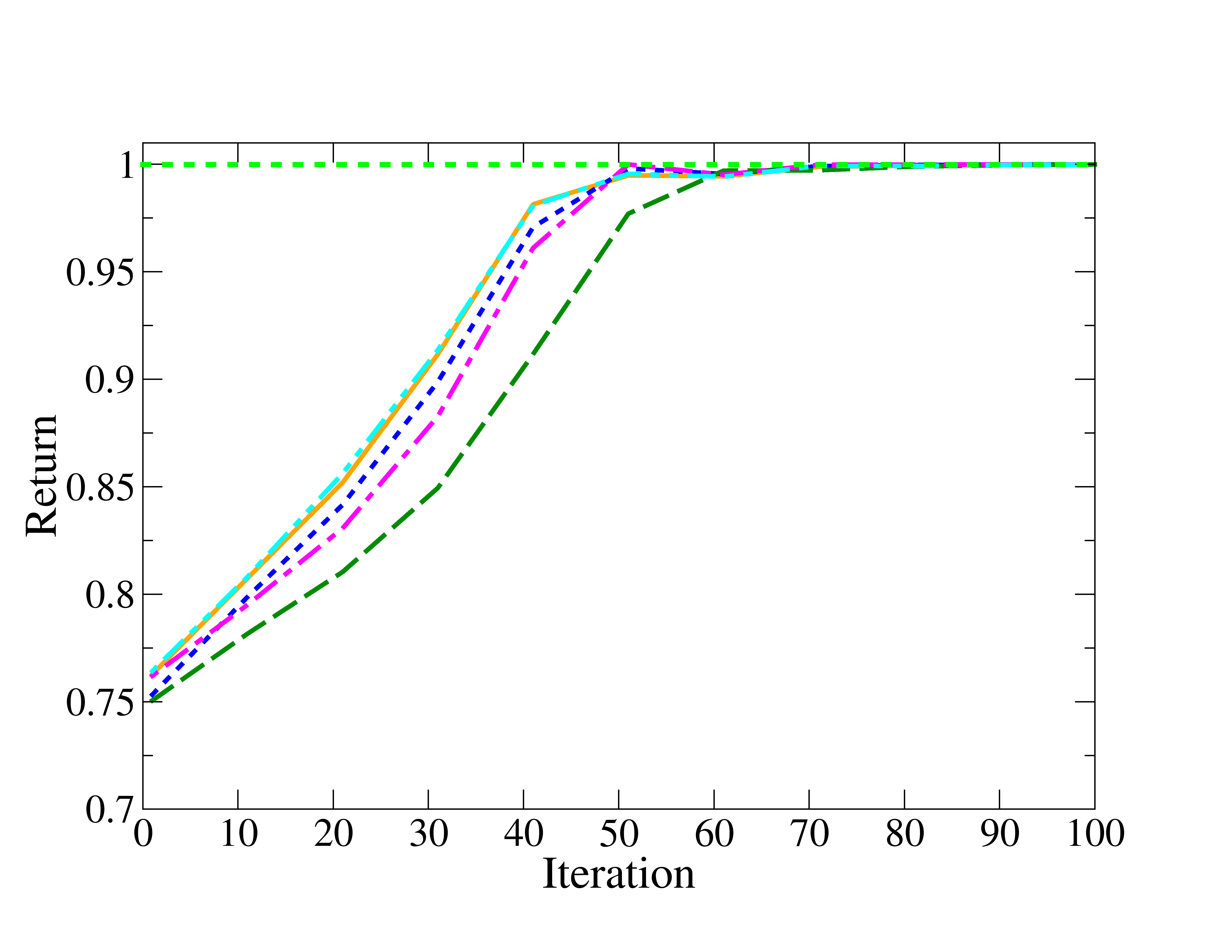}
    \caption{\justifying{Return (Eq.(\ref{eq:expectreturn})) during the optimization in the isolated system presented in figure \ref{fig:fig5} for five different simulations displayed in different colors. Each simulation runs 100 episodes. The green dotted line serves as an indicator, highlighting the target return value of 1. The random initial conditions  differ from $T\Omega = 1.5$ by about 10 percents giving a return close to 0.75.}}
    \label{fig:fig6}
\end{figure}

We first consider the ideal case without dissipation. In RL, the process begins by two Rabi frequencies chosen at random in a given range. The analytical solution (Eq.(\ref{eq:pirule})) is a landmark. By assuming a constant pulse envelope, the integrated Rabi frequency is $T\Omega$ and the best value should be $\pi / \sqrt{2}$ = 2.22. We begin the RL simulation with an initial interval with $T\Omega_{min} = 0$ and $T\Omega_{max} = 3$ that would give an area larger than the best analytical value of 2.22. Figure \ref{fig:fig5}(a) displays the first iteration with initial random Rabi frequencies, here 1.54 for $\Omega_y$ and 1.55 for $\Omega_z$. The integrated frequencies are too small and the ground state is not completely depopulated. Figure \ref{fig:fig5}(b) gives the outcome after 100 episodes. It reproduces the analytical result and is obtained after about 60 episodes as illustrated in Fig.\ref{fig:fig6}.  RL provides the good integrated frequencies but with very simple pulses since the envelopes are quasi constant. Figure \ref{fig:fig6} displays the return achieved during five simulations of 100 episodes. The random initial conditions differ from $T\Omega = 1.5$ by about 10 percents giving a return close to 0.75. Notably, the convergence rate exhibits variability and doesn't follow the typical monotonic pattern observed in conventional OCT algorithms. However, from 60 iterations, the rate achieved its highest value.

\begin{figure}
    \includegraphics[width =1.\columnwidth]{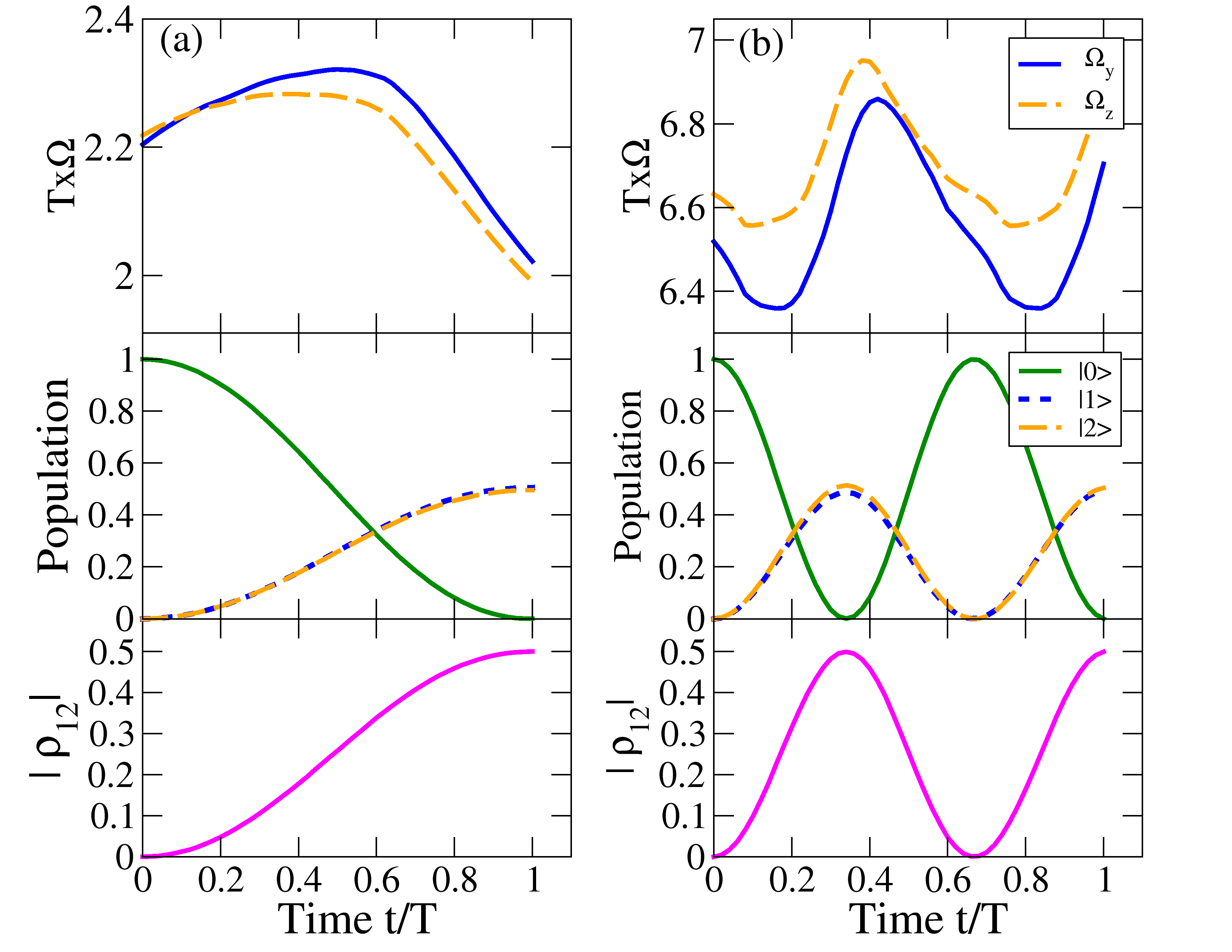}
    \caption{\justifying{Optimization of the superposed target state (Eq.(\ref{eq:target})) by RL in the isolated V-three-level system and $T\Omega_{max}=9$. The upper, middle and lower panels are as in Fig. \ref{fig:fig5}. (a) Convergence towards the best expected result ($\pi / \sqrt{2}$). The integrated frequencies are 2.286 for $\Omega_y$ and 2.254 for $\Omega_z$. (b) Convergence towards (3 $\pi / \sqrt{2}$) leading to a supplementary Rabi oscillation. The integrated frequencies are 6.673 for $\Omega_y$ and 6.825 for $\Omega_z$. } }
    \label{fig:fig7}
    \end{figure}

To test the algorithm, we start with a larger initial interval with $T\Omega_{max}= 9$. As the initial frequencies are random in this range, RL converges towards different possibilities but it is worth noting that it always finds a solution close to the analytical result. Figure \ref{fig:fig7}(a) illustrates a case where convergence occurs towards the expected value $\pi / \sqrt{2}$ with a very good final coherence. In Fig. \ref{fig:fig7}(b) one sees that according to the random initial values, optimization provides a solution with higher intensity and an area close to $3\pi / \sqrt{2}$ leading to a supplementary complete Rabi oscillation before the final coherence creation close to 0.5.  

\subsection{RL with Lindblad dynamics}
In this section, we will undertake a comparative analysis between scenarios involving constant Lindblad rates and those incorporating time-dependent variations. The constant rates associated to the four selected Lindblad operators (QuTip collapse operators) $L_k$ defined in section \ref{subsec:Lindblad} are (in reduced units $T \Gamma$): $T\Gamma_{11}= 1$ ($L_1$), $T\Gamma_{22}=0.8$ ($L_2$), $T\Gamma_{12}=0.36$ ($L_3$) and $T\Gamma_{21}=0.16$ ($L_4$). The operators $L_1$ and $L_2$ couple to the tuning baths in excited states $S_1$ and $S_2$ respectively. The ratio of the rates is approximated by $\sqrt{J_{S_2}/J_{S_1}}$ as elaborated upon in reference \cite{Jaouadi2022}. The operators $L_{3}$ and $L_{4}$  induce non-adiabatic transitions. The reduced rate $T\Gamma_{12}$ and $T\Gamma_{21}$ are different as expected from the detailed balance.  They are calibrated to roughly approximate the exact HEOM field-free dynamics at least during the early dynamics. Dynamics is performed with the QuTip mesolve solver \cite{Qutip2012}. 

\begin{figure}
    \includegraphics[width =1.\columnwidth]{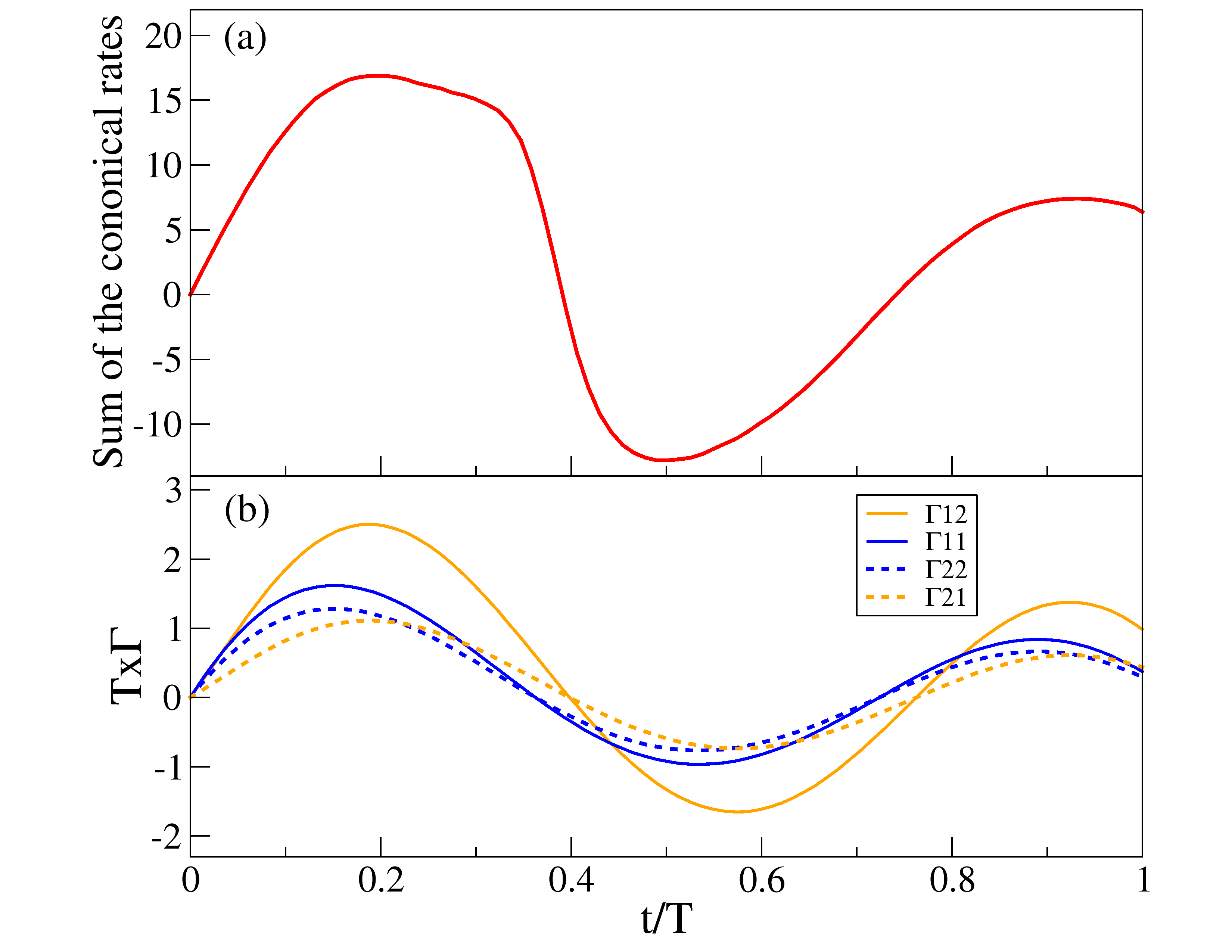}
    \caption{\justifying{(a) Sum of the canonical rates $T\Gamma^c_k$ (in reduced units) of the field-free dynamics. They are the eigenvalues of decoherence matrix (Eq.(\ref{eq:matdeco})). (b) Time-dependent rates $T\Gamma$ associated to the four Lindblad operators describing the energy tuning ($L_{1}$, $L_{2}$ with rates $\Gamma_{11}$ and $\Gamma_{22}$) and the interstate transition ($L_{3}$, $L_{4}$ with rates $\Gamma_{12}$ and $\Gamma_{21}$) induced by the coupling bath. } }
    \label{fig:fig8}
\end{figure}

Non-Markovianity may be taken into account in an approximated way by time-dependent rates. The transitory negativity of the sum of the canonical rates that are the eigenvalues of the decoherence matrix (Eq.(\ref{eq:matdeco})) is one of the signature. This sum obtained for the field-free dynamics is given in Fig.\ref{fig:fig8}(a) in reduced units ($T = 20$ fs). It is obvious that its damped oscillation follows that of the bath correlation functions (see Fig.\ref{fig:fig3}(b)). This illustrates that for this type of  partition, the non-Markovianity is closely linked to the damped vibrational motion of the collective modes. Indeed, if the collective effective mode is underdamped, the nuclei oscillate and transitory return to the initial reference position. This tends to restore the system in its initial condition. The decay towards equilibrium is not monotonous.

 Figure \ref{fig:fig8}(b) presents the time-dependent rates associated to the four Lindblad operators describing the energy tuning ($L_{1}$, $L_{2}$) and the interstate transition ($L_{3}$, $L_{4}$) induced by the coupling bath. Their shape are approximated from those of some elements of the decoherence matrix (Eq.(\ref{eq:matdeco})) by considering the basis operator $G_k$ corresponding to the $1-2$ transition (analog of $\sigma_x$ in the two-state case) and one operator corresponding to an energy gap. The amplitudes are calibrated as in the constant rate case from the field-free HEOM dynamics. The functions are fitted by polynomials or by the product of a sine function times a decreasing exponential. These functions are introduced in the time-dependent collapse operators of QuTip by using the mesolve solver \cite{Qutip2012}. 

\begin{figure}
    \includegraphics[width =1.\columnwidth]{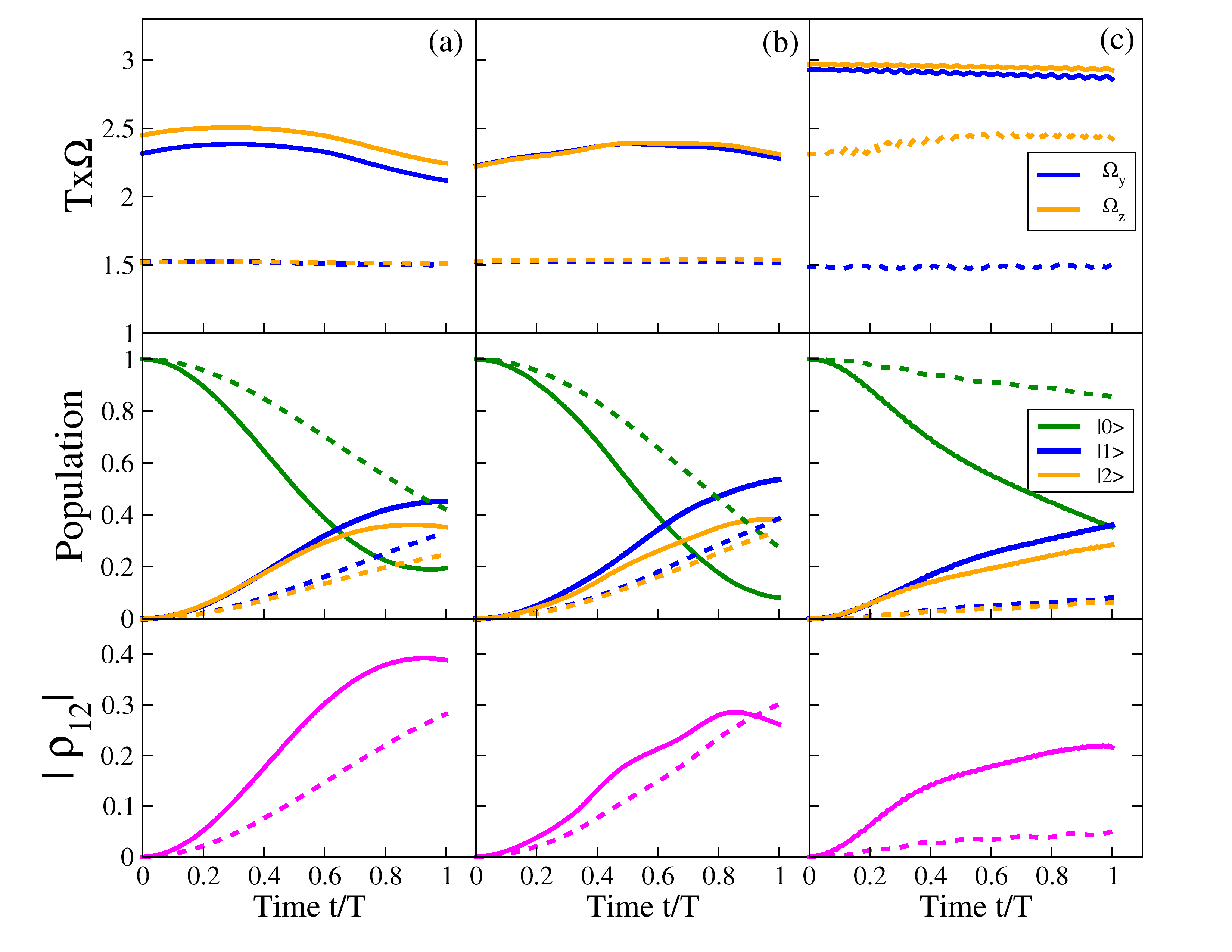}
    \caption{\justifying{Optimization of the target state (Eq.(\ref{eq:target})) by RL with Lindblad or HEOM dynamics. The upper, middle and lower panels are as in Fig. \ref{fig:fig5}. The first episode with random initial conditions is given in dashed lines}. The optimized results are given after 100 episodes. In each case, the return saturates after about 60 episodes. (a)  Dynamics with constant Lindblad rates. The area of the Rabi frequencies are 2.35 and 2.48 respectively. (b) Dynamics with time-dependent rates. The area are 2.38 and 2.39.  (c) HEOM dynamics at level 6 of the hierarchy in Sch\"odinger representation without RWA.}
    \label{fig:fig9}
    \end{figure}

 Figure \ref{fig:fig9}(a) shows the dynamics after 100 episodes with constant decay rates. The return is only 0.8 and it saturates after 60 iterations. Figure \ref{fig:fig9}(b) presents the control with the time-dependent rates. The return is slightly improved. However, this is due mainly to the better depletion of the ground state and not to a better superposition. The optimal envelopes remain very simple and quasi constant in both cases. It is a bit disconcerting that RL behaves on a very similar way with constant or time-dependent rates. We will compare these results with the OCT optimization in section \ref{sec:compareRLOCT}.

\subsection{RL with HEOM dynamics}
For a 20 fs simulation, truncating at level 6 of the hierarchy proves to be satisfactory. However, it's worth noting that for longer dynamics, such as achieving a field-free asymptotic state, a higher level 9 becomes necessary. The implementation of the RL algorithm with HEOM requires some comments. (i) During the Markov decision chain the solver is called repetitively for each  time step of the chain $\ldots {{s}_{t}}\to {{a}_{t}}\to {{r}_{t}}\to {{s}_{t+1}}\to {{a}_{t+1}}\to {{r}_{t+1}}\ldots $. All the ADOs describing the state of the surrounding must be saved for the following decision step so that each bath retains its configuration and does not restart with the initial  conditions of the baths with ADOs equal to zero. This is a difficulty that does not concern the local Lindblad dynamics. (ii) In our application, each spectral density (see Fig.\ref{fig:fig3}) is fitted by two Tannor-Meier Lorentzian functions \cite{Meier1999} leading to four decay modes for each bath. As the spectral densities are centered at high frequencies, we do not include Matsubara terms at room temperature.  We use the description of the baths by the expansion of the real and imaginary parts of the correlation functions using the BosonicBath application of the QuTip HEOMSolver \cite{Lambert2023}. When working with reduced units the real or imaginary parts of the $c_k^{R,I}$ coefficients must be scaled by $T^2$ and the rates $\gamma_k^{R,I}$ by $T$ as usually.  (iii) In the HEOM solver, the system-bath coupling operators are not written in interaction representation. Therefore, we have re-transformed the Hamiltonian in Schr\"odinger representation without the RWA approximation.  We use 100 time steps in each episode which is enough to satisfy the Nyquist-Shannon sampling rule \cite{marks2009} for the fields in the Schr\"odinger representation. 

\begin{figure}
    \includegraphics[width =1.\columnwidth]{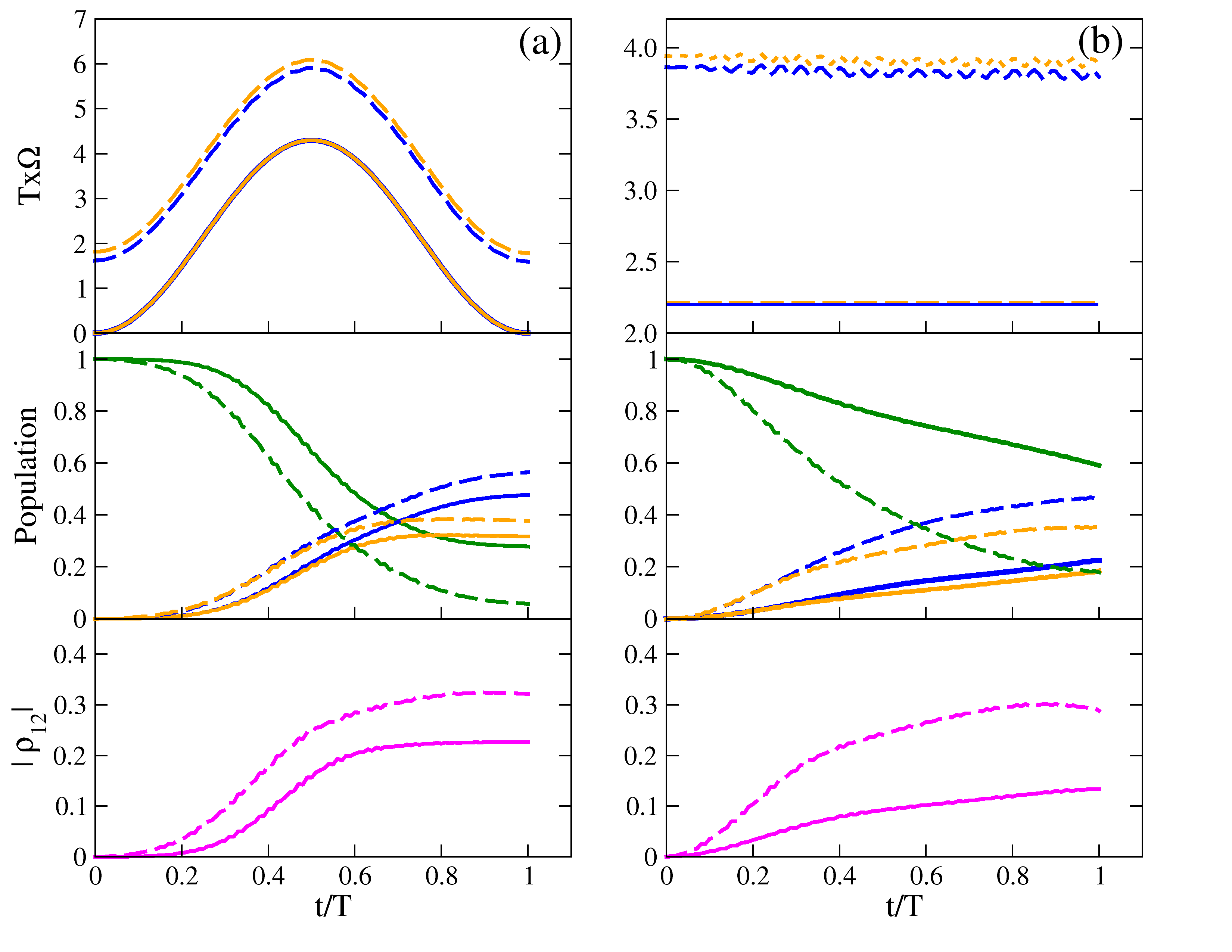}
    \caption{\justifying{RL optimization by RL with different guess fields. The actions are the variation of the Rabi frequencies $\delta \Omega$ in a range of reduced units $[-2,2]$.} Dynamics is computed by HEOM in Schr\"odinger representation without RWA at level 6 of the hierarchy. The upper panels give the Rabi frequencies $\Omega_y$ and $\Omega_z$ in reduced units, the middle ones, the populations and the lower ones, the modulus of the coherence between the two excited states. (a) guess fields with sine square envelopes of integrated Rabi frequencies $\pi/\sqrt{2}$ in solid line. The corresponding RL fields after 100 episodes in dashed line (the areas are 3.77 and 3.96). (b) guess fields with constant envelope in solid line and the corresponding RL fields after 100 episodes in dashed line (the areas are 3.87 and 3.95).}
    \label{fig:fig10}
\end{figure}

The RL optimization with HEOM in Schr\"odinger representation without RWA is given in Fig. \ref{fig:fig9}(c). The actions are more erratic due to the oscillation of the field in this representation. After 100 episodes, the envelopes have a slightly higher amplitude than in the Lindblad simulations. The populations and coherence behave on a similar way in each simulation. The return is only 0.53 primarily due to the less-than-optimal depletion of the ground state and the difference of population in the two excited states. Other examples with guess fields are given in Fig. \ref{fig:fig10}. 

Finally, we explore another strategy. We impose a guess field for the RL control by choosing the actions to be the variation $\delta \Omega$ of the Rabi frequencies with respect to the guess (see Fig.\ref{fig:fig2}(b)). These trial fields are a sine square envelope or a constant satisfying the $\pi / \sqrt{2}$ rule. Simulations are carried out in Schr\"odinger representation without RWA at level 6 of the hierarchy. The actions $T\delta \Omega$ are taken in an interval [-2,2] in reduced units. The Rabi frequencies of the guess fields and of the RL optimization after 100 episodes are given in Fig.\ref{fig:fig10}. The envelopes are only very slightly modified during the optimization. The first action is always the largest and shifts the guess envelopes by adding a constant value. The further fluctuations remain of weak amplitude. Increasing the initial interval only modifies the initial shift. Only the area increases, which generally enhances the depletion of the ground state but not the target coherence. The sine square envelope is the best guess giving a return of 80\%. The constant envelope provides only 70\%.

\section{Comparison RL-OCT}
\label{sec:compareRLOCT}

\begin{figure}
    \includegraphics[width =1.\columnwidth]{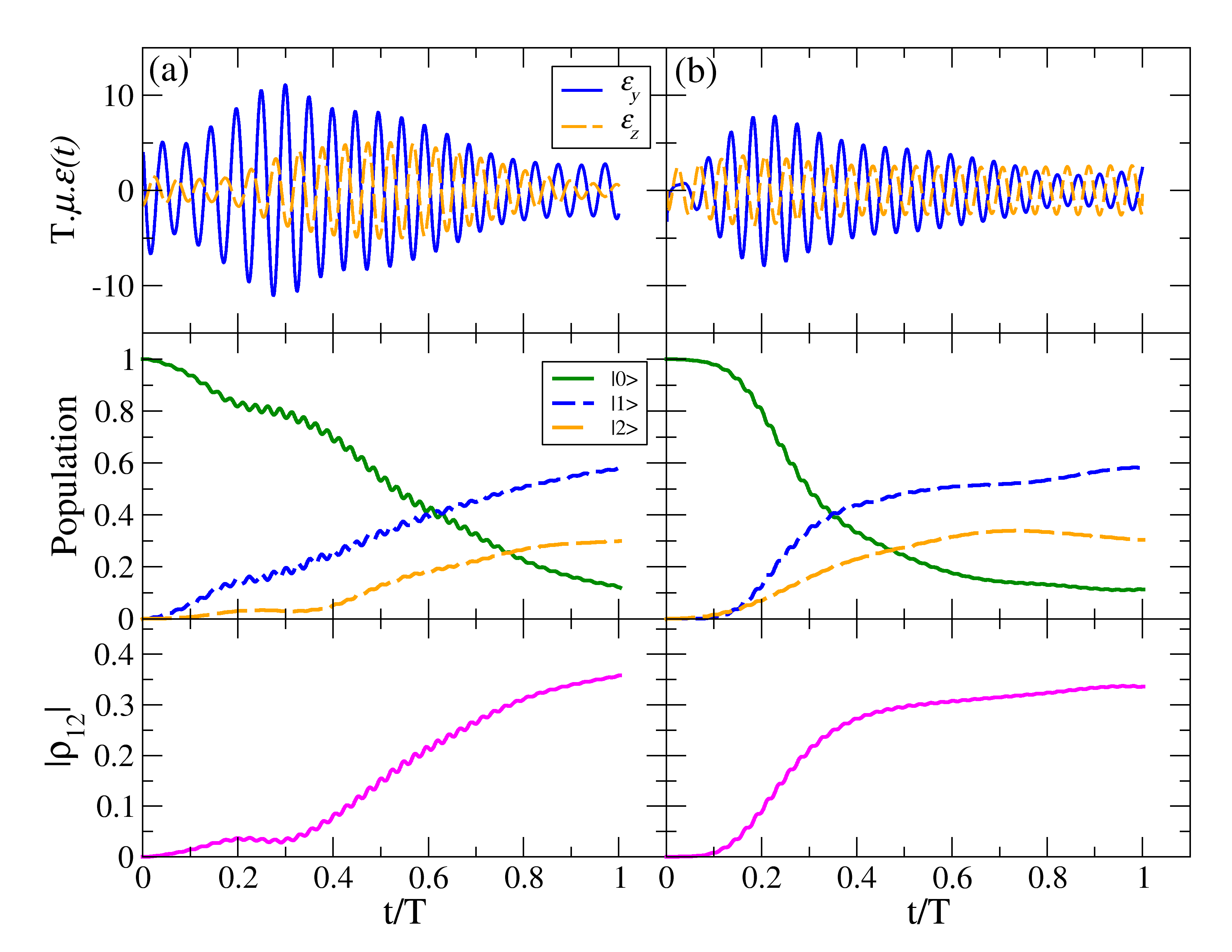}
    \caption{\justifying{Optimization by standard OCT after 15 iterations. (a) Sine square guess fields, (b) constant envelope guess fields.  The upper panels give the fields times the corresponding transition dipole in reduced units $T \mu_y\mathcal{E}_y(t)$ and $T \mu_z\mathcal{E}_z(t)$. Dynamics is computed by HEOM in Schr\"odinger representation without RWA at level 6 of hierarchy. The lower panels display the populations and the modulus of the coherence between the two excited states.}
    \label{fig:fig11}}
\end{figure}

The envelopes generated by RL consistently exhibit a high degree of simplicity, characterized by their quasi-constant profile when no guess is imposed. Given that the fields generated by standard OCT typically exhibit a higher degree of structure \cite{Jaouadi2022}, we  compare in Fig.\ref{fig:fig11} the fields optimized by OCT with the same initial guess fields drawn in solid lines in Fig.\ref{fig:fig10}. Simulations are performed utilizing our HEOM code \cite{Jaouadi2022} in Schr\"odinger representation without RWA at level 6 of the hierarchy. Standard OCT optimizes the field amplitudes and not only the envelopes on a time grid. This offers more flexibility and may slightly modify the carrier frequency. The $\alpha$ penalty factor (Eq.\ref{eq:champOCT})) is fixed to $2 \times 10^{-4}$. It influences the optimization rate. The field amplitudes increase regularly at each iteration. We take two snapshots to remain in the same order of magnitude as in the RL simulation. The results after 15 iterations are shown in Fig.\ref{fig:fig11} (a) and (b). We draw the fields times the dipole moment $T \mu_y\mathcal{E}_y(t)$ and $T \mu_z\mathcal{E}_z(t)$ in reduced units so that the envelopes may be compared with the reduced Rabi frequencies $T \Omega$ used in the RL optimization. OCT reshapes the envelopes more strongly than RL. In particular, the two envelopes do not remain similar. However, when the maximum field amplitude are in the same range, the return is similar around 80\% for the sine square case and reaches 80\% versus 70\% in RL for the constant guess.

\section{Summary and conclusion}
\label{sec:conclusion}
Examining the potential of RL in quantum control has primarily been explored within the domain of quantum information \cite{An2019,An2021,Niu2019,Prati2021}. This analysis is particularly interesting in the context of retrieving the STIRAP scheme using either digital pulses \cite{Prati2020} or continuous ones \cite{Brown2021,Giannelli2022}. The outstanding property is the ability to propose strategies without any prior knowledge of the system leading to the denomination as a "model-free algorithm" \cite{Marquardt2021}. The main question is to see whether RL will find new strategies in particular in presence of an environment. Most of the previous works have studied examples with dissipation treated by Lindblad master equation, i.e. for a Markovian dynamics \cite{An2021,Giannelli2022}. However, even if RL is built on a succession of decision Markov processes, non-Markovian noise could influence the system dynamics \cite{Porotti2019}. 

In this work, we have revisited a control in a system with a strong non-Markovian dynamics due to the coupling to baths with highly structured spectral densities leading to long bath correlation times. The model is calibrated from $\textit{ab}$ $\textit{initio}$ data \cite{Ho2017,Jaouadi2022,Lasorne2023}. We have used an open source software \cite{Giannelli2022,RLpy} based on the policy gradient method for the optimization and on QuTip mesolve solver of the Lindblad master equation \cite{Qutip2012}. We have enhanced its functionality to address non-Markovian dynamics. In a first approximate go-between step we have incorporated time-dependent rate constants derived from the field-free HEOM dynamics. Then we have interfaced the RL algorithm with the HEOM solver of the QuTip BoFiN package \cite{Lambert2023}.

An analytical solution exists to create the target superposition of two excited states addressable with orthogonal dipoles in an isolated V-system. It is of significant interest to assess the proficiency of RL to recover the expected solution from random initial conditions in a given interval for the Rabi frequencies. RL finds a very simple but efficient solution of straightforward quasi-constant envelopes satisfying the integrated Rabi frequency rule. The amplitudes display minimal variations, typically within a few percentage points. 

For each level of complexity of the dynamics, we observe that the return saturates after about 60 iterations even if the target is not perfectly reached. The Rabi frequencies are always very simple, nearly constant when no guess is imposed. The proposal is basic and robust. Even if it is not completely satisfactory, RL does not go further. When dynamics is driven in Schr\"odinger representation without RWA, the process of optimizing using reinforcement learning (RL) exhibits increased complexity. This complexity is reflected in the erratic behavior of the envelopes from one step to another. However, it's worth noting that despite these fluctuations, a smoother average trajectory is observed with only minor fluctuations. 

Lindblad dynamics even with some time-dependent rates cannot take into account a possible influence of the field on the baths. On the contrary, the memory kernel of HEOM contains the time dependent Hamiltonian \cite{Meier1999} and this could in principle induce an effect on the bath dynamics \cite{Chenel2012}. Indeed, the decoherence matrix and thus the rates are modified by the field \cite{MDL2018}. However, in our application the behavior is qualitatively the same for the approximate non-Markovian approach or for exact HEOM. RL successfully captures the crucial condition regarding the integrated Rabi frequency, yet it does not discover novel strategies to fight dissipation. It would be powerful to increase the number of available actions, enabling optimization of detuning parameters as well or directly the amplitude of the fields and not only the envelopes. Another possibility would be to let the algorithm choose a guess field. Furthermore, the exploration of more sophisticated RL algorithms holds promise for future investigations \cite{Reuer2022,Schulman2017}. 
 
 The simplicity of the RL envelopes suggests to confront the standard OCT and to see if it can yield superior results. OCT exploits the system dynamics and may seem more flexible since it optimizes the field amplitude and the carrier frequency and possibly finds some chirp effect. However, in our example OCT is not more efficient to reach the target with dissipation. By imposing the same guess fields, RL and OCT provide different optimal fields ensuring similar return. The reshaping is stronger in OCT that proposes different envelopes for the two polarizations what RL does not do. When the envelope amplitudes are maintained in the same range as in RL, OCT slighly improves the depletion of the ground state but not really the preparation of the superposition with equal weights. Both strategies, RL and OCT depend on the guess fields and the optimal fields are different. However, they ensure similar final dynamics and the perfect target is not achieved neither by RL nor by OCT control due to the strong dissipation.  

 Our example is a complex system strongly coupled to a structured environment with laser pulses in the femtosecond range. RL seems more adapted to treat quantum information in another spectral range operating with very simple square box envelopes and weakly coupled Markovian noises\cite{Baum}.

\section{Data availability}
The data are available upon request to the authors. The modified ThreeLS.py file of Luigi Giannelli's open-source software \cite{Giannelli2022,RLpy} allowing dynamics with HEOM in Schr\"odinger representation without RWA by using the QuTip BoFiN package \cite{Lambert2023} is given in supplementary material \cite{SM}. 

\section{Acknowledgements}
The authors thanks Dr O. Atabek for stimulating discussions and encouragements to explore this project.  Dr B. Lasorne and J. Galiana from ICGM, Univ Montpellier, CNRS, ENSCM, Montpellier, France, are acknowledged for providing the $\textit{ab}$ $\textit{initio}$ date calibrating the model. 

\bibliography{bibRL}

\end{document}


\title{Supplementary Material \\
 Re-exploring Control Strategies in a Non-Markovian Open Quantum System by Reinforcement Learning}

\author{Amine Jaouadi}
 \affiliation{LYRIDS, ECE-Paris, Graduate School of Engineering, 75015 Paris, France} 

\author{Etienne Mangaud}
\affiliation{MSME, Universit\'e Gustave Eiffel, UPEC, CNRS, F-77454 Marne-La-Vall\'ee, France }
 
\author{Mich\`ele Desouter-Lecomte}
\affiliation{Institut de Chimie Physique, Universit\'e Paris-Saclay-CNRS, UMR8000, F-91400 Orsay, France}

\begin{abstract}
The supplementary material gives the modification of the ThreeLS.py file of Luigi Giannelli's open-source software \cite{b1,b2} allowing dynamics with HEOM in Schr\"odinger representation without RWA by using the QuTip BoFiN package \cite{b3}.
\end{abstract}
\maketitle

\section{Import Libraries}
\begin{verbatim}
import pickle
import matplotlib.pyplot as plt
import numpy as np
# using the style for the plot
plt.style.use('seaborn-poster')
from qutip import *
from qutip.nonmarkov.heom import DrudeLorentzBath
from qutip.nonmarkov.heom import DrudeLorentzPadeBath
from qutip.nonmarkov.heom import BosonicBath
from qutip.nonmarkov.heom import HEOMSolver
from qutip.nonmarkov.heom import HierarchyADOs

from scipy.optimize import minimize
from tf_agents.environments import py_environment
from tf_agents.specs import array_spec
from tf_agents.trajectories import time_step as ts
from tqdm.auto import trange
\end{verbatim}

\section{Initialize the system}
\begin{verbatim}
opts = Options(atol=1e-11, rtol=1e-9, nsteps=int(1e6))
opts.normalize_output = False  # mesolve is x3 faster if this is False
class ThreeLS_v0_env(py_environment.PyEnvironment):
    """Lambda-system environment"""

    def __init__(
        self,
        Omax=3.0,
        Omin=0.0,
        Delta=0.16277*840,     # energies in ua 
                               #scaling factor T (840 = 20fs)
        delta_p=0.16439*840, 
        T=1.0,
        n_steps=80,
        Gamma=0.0,
        reward_gain=1.0,
        seed=1,
        max_depth = 6,        
    ):
        
        self.qstate = [basis(4, i) for i in range(4)]
        self.sig = [
            [self.qstate[i] * self.qstate[j].dag() for j in range(4)] for i in range(4) ]
        self.up = (self.sig[0][1] + self.sig[1][0]) 
        self.us = (self.sig[0][2] + self.sig[2][0]) 
        self.Psi_0 = ket2dm(self.qstate[0])
        self.target_state=(self.sig[2][2]+self.sig[1][1]+self.sig[1][2]+self.sig[2][1])/2

        self.Omax = Omax
        self.Omin = Omin
        self.Delta = Delta 
        self.delta_p = delta_p
        self.omegay = Delta 
        self.omegaz = delta_p 
        self.T = T
        self.n_steps = n_steps
        self.Gamma = Gamma
        self.reward_gain = reward_gain
        self.current_step = 0
        self.current_qstate = self.Psi_0
        self._state = self._dm2state(self.Psi_0)
        self._episode_ended = False
\end{verbatim}

\section{Bath properties}  
\begin{verbatim}
        self.Temp = 0.001  # Temp not used when the correlation
                           #function is given in in the Bosonic Bath      
        # System-bath coupling operator:        
        self.Qc = 1. * (self.sig[1][2]+self.sig[2][1])
        self.Qt = self.sig[1][1] + 0.89 * self.sig[2][2]

        # Matsubara expansion:
        nmod = 4    # number of decay modes for each bath
        fac = 840.  # scaling factor for the frequancies
        fac1 = fac*840.  # scaling factor the coefficients 
                         # of the expansion of the correlation function

    # Enter real and imaginary parts of the coefficients for each bath        
      file=open('ckrt',"r")
        self.cktr=[]
        for i in range(nmod):
           line = file.readline().split() 
           self.cktr.append(complex(float(line[0]),float(line[1])) * fac1)
        
        file=open('ckit',"r")
        self.ckti=[]
        for i in range(nmod):
           line = file.readline().split()
           self.ckti.append(complex(float(line[0]),float(line[1])) * fac1)
           
        file=open('ckrc',"r")
        self.ckcr=[]
        for i in range(nmod):
           line = file.readline().split()
           self.ckcr.append(complex(float(line[0]),float(line[1])) * fac1)   
        
        file=open('ckic',"r")
        self.ckci=[]
        for i in range(nmod):
           line = file.readline().split()
           self.ckci.append(complex(float(line[0]),float(line[1])) * fac1)

    # Enter real and imaginary parts of the frequancies for each bath
        file=open('nukrt',"r")
        self.nuktr=[]
        for i in range(nmod):
           line = file.readline().split()
           self.nuktr.append(complex(float(line[0]),float(line[1])) * fac)
        
        file=open('nukit',"r")
        self.nukti=[]
        for i in range(nmod):
           line = file.readline().split()
           self.nukti.append(complex(float(line[0]),float(line[1])) * fac)
        
        file=open('nukrc',"r")
        self.nukcr=[]
        for i in range(nmod):
           line = file.readline().split()
           self.nukcr.append(complex(float(line[0]),float(line[1])) * fac)
        
        file=open('nukic',"r")
        self.nukci=[]
        for i in range(nmod):
           line = file.readline().split()
           self.nukci.append(complex(float(line[0]),float(line[1])) * fac)
           
        self.batht = BosonicBath(self.Qt,self.cktr,self.nuktr,self.ckti,self.nukti)
        self.bathc = BosonicBath(self.Qc,self.ckcr,self.nukcr,self.ckci,self.nukci)
        self.bath = [self.batht,self.bathc]
        #nheom
        self.max_depth = 6  # maximum hierarchy depth to retain
        self.ados = HierarchyADOs(
            self._combine_bath_exponents(self.bath),
            self.max_depth,
        )
        self.update()

    def _combine_bath_exponents(self, bath):
        """ Combine the exponents for the specified baths. """
        if not isinstance(bath, (list, tuple)):
            exponents = bath.exponents
        else:
            exponents = []
            for b in bath:
                exponents.extend(b.exponents)
        all_bosonic = all(exp.type in (exp.types.R, exp.types.I, exp.types.RI)
                          for exp in exponents)
        all_fermionic = all(exp.type in (exp.types["+"], exp.types["-"])
                            for exp in exponents)
        if not (all_bosonic or all_fermionic):
            raise ValueError(
                "Bath exponents are currently restricted to being either"
                " all bosonic or all fermionic, but a mixture of bath"
                " exponents was given.")
        if not all(exp.Q.dims == exponents[0].Q.dims for exp in exponents):
            raise ValueError(
                "All bath exponents must have system coupling operators"
                " with the same dimensions but a mixture of dimensions"
                " was given.")
        return exponents    
    \end{verbatim}
    
  \section{Modification of mesolvestep function FOR HEOM}
  \begin{verbatim}
    def _mesolvestep(self, action, qstate):
        tlist = self.tlist[self.current_step : self.current_step + 2]                  
        fp = function_from_array((action[0])*(np.cos(self.omegay*tlist)), tlist)
        fs = function_from_array((action[1])*np.cos(self.omegaz*tlist), tlist)
        H = [self.H0, [self.up, fp], [self.us, fs]]
        solver  = HEOMSolver(H, self.bath, max_depth=self.max_depth, options=opts)
        if self.current_step == 0:
           result = solver.run(qstate, tlist,ado_return=True)
           self.ados=result.ado_states[-1]
           return result.states[-1]
        else:
          result = solver.run(self.ados, tlist,ado_init=True,ado_return=True)
          self.ados=result.ado_states[-1]
          result.states[-1] = result.ado_states[0].extract(0)
          return result.states[-1]
    \end{verbatim}
\section{Modification of run mesolveolution function FOR HEOM}
   \begin{verbatim}
    def run_mesolvevolution(self, amps):
        Omega_p = amps[:, 0]
        Omega_s = amps[:, 1]
        fp = function_from_array((Omega_p)*np.cos(self.omegay*self.tlist[:-1]),
              self.tlist[:-1])
        fs = function_from_array((Omega_s)*np.cos(self.omegaz*self.tlist[:-1]), 
              self.tlist[:-1])
        H = [self.H0, [self.up, fp], [self.us, fs]]
        solver = HEOMSolver(H, self.bath, max_depth=self.max_depth, options=opts)
        result = solver.run(qstate, self.tlist)
        return result.states[-1]
\end{verbatim}